# 38.7 GHz Thin Film Lithium Niobate Acoustic Filter

Omar Barrera, Sinwoo Cho, Jack Kramer, Vakhtang Chulukhadze, Joshua Campbell and Ruochen Lu
The University of Texas at Austin, Austin, TX, USA
omarb@utexas.edu

*Abstract*—In this work, a 38.7 GHz acoustic wave ladder filter exhibiting insertion loss (IL) of 5.63 dB and 3-dB fractional bandwidth (FBW) of 17.6% is demonstrated, pushing the frequency limits of thin-film piezoelectric acoustic filter technology. The filter achieves operating frequency up to 5G millimeter wave (mmWave) frequency range 2 (FR2) bands, by thinning thin-film LiNbO$_3$ resonators to sub-50 nm thickness. The high electromechanical coupling ($k^2$) and quality factor ($Q$) of first-order antisymmetric (A1) mode resonators in 128° Y-cut lithium niobate (LiNbO$_3$) collectively enable the first acoustic filters at mmWave. The key design consideration of electromagnetic (EM) resonances in interdigitated transducers (IDT) is addressed and mitigated. These results indicate that thin-film piezoelectric resonators could be pushed to 5G FR2 bands. Further performance enhancement and frequency scaling calls for better resonator technologies and EM-acoustic filter co-design.

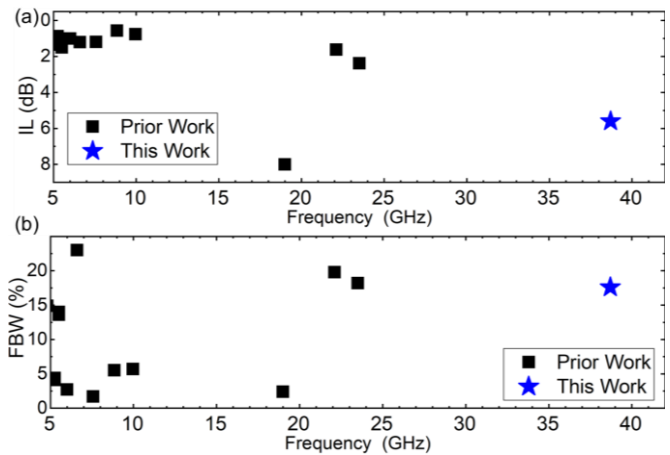

Fig. 1 Survey of (a) IL and (b) FBW in acoustic filters above 5 GHz.

## I. INTRODUCTION

The next generation of wireless network communications requires devices working at increasingly higher frequencies, e.g., millimeter wave bands like new radio (NR) frequency range 2 (FR2) channels, to cater to the demand for large data rates [1]–[3]. In particular, filter technologies for mobile devices need not only to operate at higher bands, but they must maintain small footprints [4]. Acoustic resonator-based filters have traditionally fulfilled this role since acoustic wave's shorter wavelengths than electromagnetic (EM) counterparts allow device miniaturization [5], [6]. However, commercially available acoustic filters have remained stagnant at sub-6 GHz [7] frequency bands, as scaling resonant frequency of existing acoustic resonator technologies, e.g., conventional film bulk acoustic wave (FBAR) and surface acoustic wave (SAW), has proven to be a formidable challenge. The most straightforward techniques involve very precise lithography techniques [8]–[10], or very thin piezoelectric layers [11], methods which are always limited by the available technology.

Nevertheless, recent academic progress has been made towards realizing high performing high-frequency acoustic resonators by harnessing/utilizing promising materials such as lithium niobate (LiNbO$_3$) [12]–[14] and aluminum nitride/scandium aluminum nitride (AlN, ScAlN) [15], [16]. Utilizing such high-performing materials with innovative new design approaches has resulted in the first-order antisymmetric (A1) [17], [18] mode, as well as overmoded acoustic resonators in piezoelectric thin films. The advantages of such an approach for frequency scaling become clear since the frequency of operation is largely dictated by the layer thickness, provided it is much smaller than the lateral dimensions, relaxing the feature size requirement for lateral dimensions that are lithographically defined [19]. Thus, in the case of thin-film A1 devices [20], [21] the only theoretical limit to scaling frequency is the thickness of the film. Recent reports have shed light on the ability to trim the thickness of acoustic resonators using controlled ion milling, the results indicate good preservation of the film quality and the figure of merit (FOM) of the resonators [22]. Employing such techniques has resulted in groundbreaking 22 GHz filters in LiNbO$_3$ resonators working at the A1 mode [23]. Despite the advance, such demonstrations are still below the 5G FR2 bands. The challenges and opportunities in further frequency scaling LiNbO$_3$ technology remain unexplored.

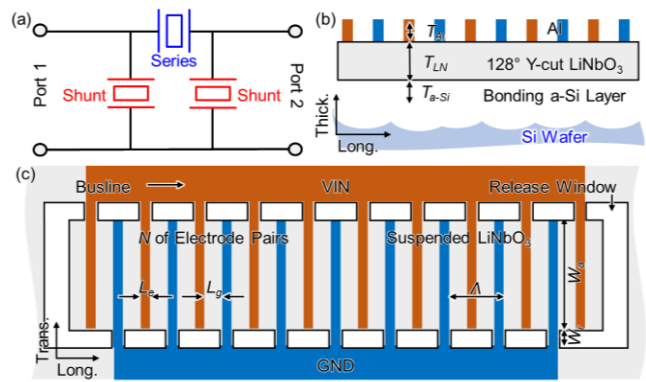

Fig. 2 (a) Filter circuit, and resonator (b) top and (c) side schematics.

In this paper, a 38.7 GHz acoustic filter on thin film LiNbO3 is reported. The filter achieves an insertion loss (IL) of 5.63 dB, out-of-band rejection of 15.8 dB, and a fractional bandwidth (FBW) of 17.6%. These results mark the highest frequency response ever reported in acoustic wave filters. Moreover, the electromagnetic resonances that naturally occur in interdigitated transducers (IDT) are addressed and kept outside the filter passband. In doing so, this work theorizes that future IDT-based acoustic resonator filters working beyond 40 GHz are realizable, given further resonator performance enhancement and a more thorough EM-acoustic filter co-design. Thus future-generation acoustic filter designers will have to either employ EM resonances to their advantage during the synthesis or look for ways to reject them.

## II. DESIGN AND SIMULATION

The designed device is a third-order ladder filter with 2 shunt resonators, one at each port, and one series resonator as

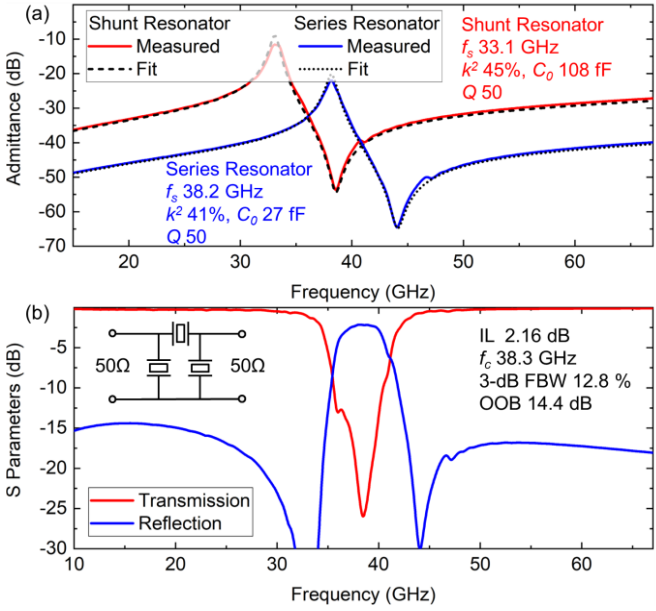

Fig. 3 (a) Simulated resonator admittance amplitude and extract key resonator performance. (b) Simulated filter response with 50 Ω terminations.

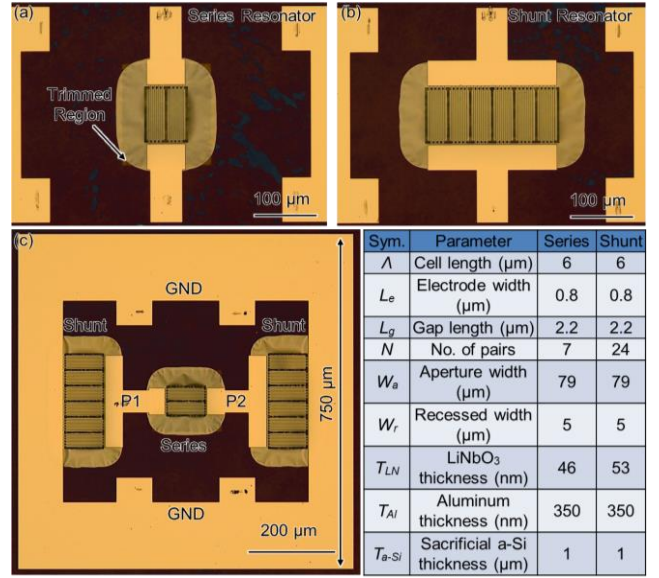

Fig. 4 Microscopic images of fabricated (a) series resonator, (b) shunt resonator, and (c) filter. Key dimensions are listed in the table.

depicted in Fig 2 (a). The resonators are laterally excited devices using IDT's to excite $e_{15}$ of a 128° Y-cut LiNbO$_3$ piezoelectric layer. The full material stack consists of thin-film LiNbO$_3$ on over 1 μm amorphous silicon (a-Si), used for bonding and sacrificial purposes, on top of a 500 μm silicon carrier substrate. The spacing between adjacent IDT's is 3 μm and aluminum is selected as the metal layer since it has good selectivity during silicon etching and device release. More details on the device dimensions are listed in Fig 2. (c). The resonator design is largely based on the platform proposed in [20]. The resonator performance is first evaluated using COMSOL finite element analysis (FEA), the simulated and results for the A1 mode are plotted in Fig. 3(a). The electromechanical coupling ($k^2$) is extracted using the modified Butterworth-Van Dyke (MBVD) fitting, approximated by $k^2 = \pi^2/8 \cdot (f_p^2/f_s^2 - 1)$. The model does not include the EM resonances expected to occur around 50 GHz, and the associated frequency shifts it incurs on the resonator, experimentally observed in our earlier work on 22 GHz filters [23]. The results show series resonances ($f_s$) at 33 and 38 GHz, with $k^2$ of 45 and 41% for the shunt and series resonators respectively, very well suited for the FR2 band. The difference in $k^2$ arise from the thickness variation between resonators, 53 nm shunt and 46 nm series, required to synthesize the ladder filter. The static capacitances are chosen so as to achieve minimum IL with 50Ω port impedances. The simulated FEA data is exported and used to synthesize and simulate the filter with the aid of Cadence AWR. The results of the simulation [Fig. 3 (b)] show passband centered at 38.3 GHz, with an IL of 2.16 dB, 12.8 % FBW and out-of-band rejection of 14.4 dB.

III. FABRICATION AND MEASUREMENT

The devices are fabricated on LiNbO$_3$-aSi-Si wafers provided by NGK Insulators Ltd., the specification thickness of the LiNbO$_3$ layer is 110 nm. As a first step the sample (2.1 by 1.9 cm) is trimmed down using an ion milling etch process to 53 nm to provide the base thickness required by the shunt resonators. The desired thickness is monitored using a Woollam ellipsometer. Next, etching windows are patterned using lithography. Long lateral etch windows are allocated along the resonators length, effectively dividing them into resonator banks, which helps the release process. A second round of ion milling is performed to etch the LiNbO$_3$ thin-film through the etch windows, deep into the a-Si layer. Afterwards, local trimming regions are lithographically defined, and a third round of ion milling is carried on to further trim the thickness to these regions to the target 46 nm. The local trimmed regions are used for the series resonators, and they accomplish the frequency shift required to realize the filter. The step height difference between the base thickness and the local trimmed regions is verified using atomic force microscopy (AFM), and measured as 8 ±1 nm at different regions on the sample. The aluminum electrodes are patterned using electron beam lithography (EBL) and metal evaporation. Finally, the resonators are released using xeon difluoride (XeF$_2$) for silicon selective etching. Optical pictures of the filter and the standalone resonators are shown in Fig. 4 (a), (b) and (c). Detailed dimensions are listed in the table included in Fig. 5. The filter layout is based on the one reported in [23].

The performance of the resonators and filters is evaluated using a Keysight vector network analyzer (VNA) at a fixed -10 dBm power level. The admittance and phase response of the standalone resonators are displayed in Fig. 5 (a) and (b), along with detailed extracted MBVD parameters in (c). The routing components $L_s$ and $R_s$ are required to fit the EM resonance embedded in the structure. The measurements show a small discrepancy for the perceived $f_s$ of the series resonator, which is likely caused by a slight over etch of the local region and the interaction with the EM resonance at 50 GHz. The resonators show $k^2$ of 30 with quality factor $Q$ of 13 for the shunt resonator, and $k^2$ of 25 with $Q$ of 10 for the series resonator.

The filter measurements under 50 Ω [Fig. 6(a)] exhibit a passband centered at 38.7 GHz with IL of 5.63 dB, broad 3-dB FBW of 17.6 % and OoB rejection of 15.8 dB. The filter is the first demonstration of the feasibility of using acoustic technologies for mmWave filters [Fig. 1 (a)(b)].

To further analyze loss and parasitics in the filter, the measurements show impedance mismatch from the return loss, deviated from the design. This is due to a combination of

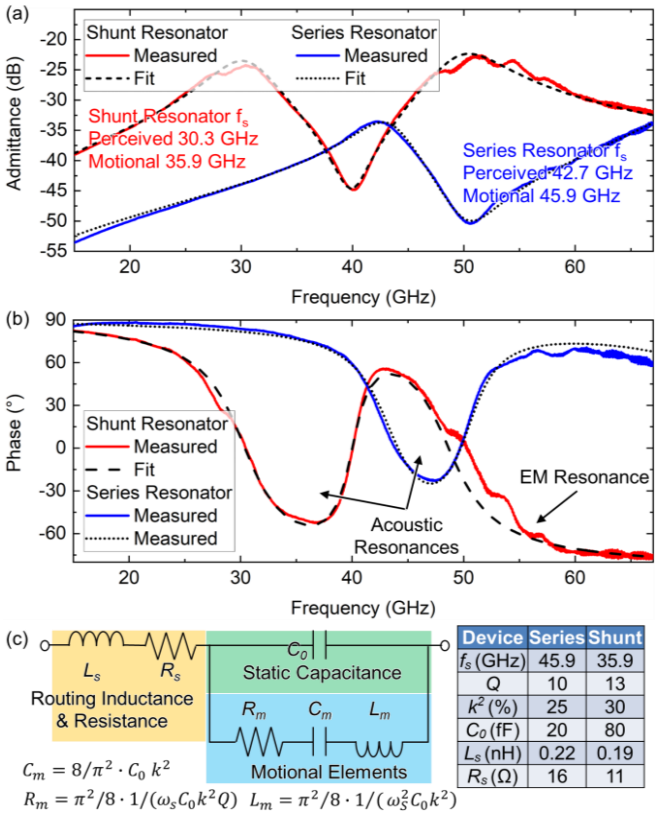

Fig. 5 Measured wideband admittance response in (a) amplitude and (b) phase. (c) Modified mmWave MBVD model and extracted key resonator specifications.

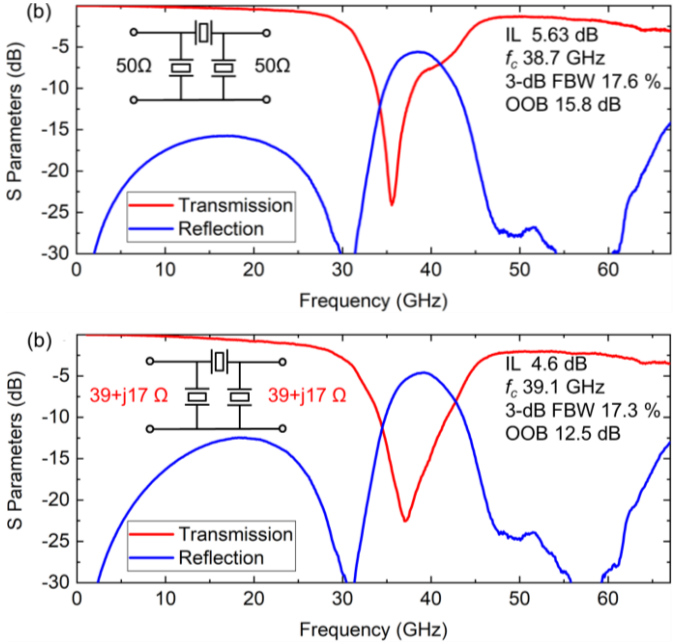

Fig. 6 (a) Measured raw filter data transmission and reflection with 50 Ω. (b) Filter transmission and reflection using complex matched source and loads with 39+j17 Ω.

factors, including the discrepancy in perceived frequency response of the resonators, the effects of the EM resonances as well as higher frequency parasitics embedded in the filter layout. The measurement data is then artificially impedance matched in AWR using complex input and output sourced impedances to predict the filter performance under matched conditions. The results are shown in Fig. 6 (b) with 39+j17 Ω port impedance. The IL of 4.6 dB, now eliminating the port reflection, is the loss in the filter, collectively contributed by the electrical routing loss, and mechanical loss. The loss can be further reduced by a series of methods, including using multiple-layer piezoelectric with alternating orientations, so-called periodically poled piezoelectric film (P3F) [24]–[26] to operate at higher-order Lamb modes, thickening up buslines for less routing resistive loss, and EM-acoustic co-designs for mitigating the inductive parasitics.

IV. CONCLUSION

This work reports a thin-film $LiNbO_3$ acoustic filter at 38.7 GHz with an IL of 5.63 dB and a 3-dB FBW of 17.6 %. The results are the highest in frequency response ever reported and demonstrate the realizability of acoustic wave filters for next-generation radio bands.

ACKNOWLEDGMENT

The authors thank the DARPA COFFEE program for funding support and Dr. Ben Griffin for helpful discussions.